\documentclass[]{interact}

\usepackage{epstopdf}% To incorporate .eps illustrations using PDFLaTeX, etc.
\usepackage[caption=false]{subfig}% Support for small, `sub' figures and tables
\usepackage{cite}
\usepackage{geometry}
\usepackage{cancel}
\usepackage{algorithm}
\usepackage{algpseudocode}

\usepackage{mathtools}

\usepackage{amsmath}
\usepackage{amssymb}
\usepackage{mathtools}
\usepackage[mathscr]{euscript}
\usepackage[frozencache,cachedir=minted-cache]{minted}
\usepackage{listings}
\usepackage{forest}
\usepackage{newclude}
\usepackage{diagbox}
\usepackage[numbers,sort&compress]{natbib}% Citation support using natbib.sty
\bibpunct[, ]{[}{]}{,}{n}{,}{,}% Citation support using natbib.sty
\renewcommand\bibfont{\fontsize{10}{12}\selectfont}% Bibliography support using natbib.sty
\theoremstyle{plain}% Theorem-like structures provided by amsthm.sty

\theoremstyle{definition}

\theoremstyle{remark}

\newcommand{\mult}{\text{\hspace{0.1cm}}}
\DeclarePairedDelimiter\abs{\lvert}{\rvert}%
\begin{document}
%\articletype{ARTICLE TEMPLATE}% Specify the article type or omit as appropriate
\title{Differentiable Programming: Efficient Smoothing of
	Control-Flow-Induced Discontinuities}

\author{
\name{Sebastian Christodoulou\thanks{Contact S. Christodoulou. Email: christodoulou@stce.rwth-aachen.de} and Uwe Naumann}
\affil{
	%\textsuperscript{a}
	Software Tools for Computational Engineering, RWTH Aachen University, Germany%; 
	%\textsuperscript{b}Institut f\"{u}r Informatik, Albert-Ludwigs-Universit\"{a}t, Freiburg, Germany
	}
}

\maketitle

\begin{abstract}
We want to obtain derivatives in discontinuous program code, where default Algorithmic Differentiation may not perform well. Specifically, we consider discontinuities induced by control flow statements, where meaningful derivatives should ideally be capable of representing the resulting 'jumps' in the trajectory. To achieve this, one can interpolate the trajectory at the control flow statements before taking the derivative. We formulate a method to efficiently interpolate between all boundaries induced by control flow in program code. Theoretically, code can be viewed as a series of piecewise continuous functions applied in succession. These functions are nested inside one another and result in a function composition with several cases. We interpret this function composition as a tree and devise a heuristic to identify paths that are relevant to the interpolation. This allows us to conceive a language that smoothly interpolates control-flow statements automatically and efficiently, making it fully differentiable. 
\end{abstract} 

% AVOID LINE-BREAK:
%\include*{properties}
\section{Introduction}
Differentiated program code has wide-ranging applications, from stability estimates for solvers to sensitivities in computational finance \cite{capriotti2010fast}. We are interested functions that are piecewise continuous and can be formulated as a computer program. Examples include functions with discontinuous payoffs or costs \cite{siegel2014asymmetric} and piecewise continuous models. In principle, Algorithmic Differentiation (AD) can be deployed on these programs, but might not result in meaningful derivatives at discontinuities: Generated derivatives simply switch by the same rules in the control flow as the primals do. This limits the utility of AD in discontinuous settings \cite{more2014you}. We want to generate derivatives that account for the 'sudden jump' caused by control flow and for the trajectory beyond it. We develop a language, where control flow statements are fully differentiable.
% By limiting ourselves to discontinuities caused by control flow, we use can use the adjoint mode of AD [] for derivatives, while retaining the advantage which is otherwise unique to finite differences: The representation of the derivative's trajectory despite discontinuities [].  %Furthermore, we introduce a parameter to tweak the sharpness of this interpolation. \\

%%%%%%%%%%%%%%%%%%%%%%%%%%%%%%

Our method draws inspiration from the idea that a discontinuity can be replaced by interpolating in between the cases of a piecewise-defined function. The idea to use interpolation in this way to obtain meaningful derivatives goes back to Bertsekas \cite{bertsekas1975nondifferentiable} for nondifferentiabilities, and Zang \cite{zang1981discontinuous} for piecewise continuous functions. Along with similar work undertaken in this direction more recently \cite{chakraborty2016non, reparamGradientSmooth2023}, approaches generally rely on \emph{weighing} all of a piecewise continuous function's cases to interpolate between them: Each case is evaluated for a given input, where a weight determines how closely it applies.
%%%%%%%%%%%%%%%%%%%%%%%%%%%%%%
%Our method draws inspiration from the idea that a discontinuity can be replaced by interpolating in between the cases of a piecewise-defined function. Derivatives are then obtained from the interpolated function. Some work has been done in this direction \cite{zang1981discontinuous, reparamGradientSmooth2023}. Roughly speaking, existing approaches weigh all cases of a discontinuous function by \emph{how closely} those cases apply to a given input. 

However, viewing program code through this lens is not always beneficial. Even though a program's paths through the control flow can be mapped to cases, decisions and computations are generally intertwined within the control flow. Consequently, to know whether any given case applies, most of the associated program execution has to take place beforehand. 

A more general approach to obtaining smooth derivatives of any discontinuous or noisy function is to regularize the differentiation process based on available function evaluations \cite{chartrand2011numerical}. Sample-based approaches are also popular in nonsmooth optimization, where gradients around the point of interest are sampled and then interpolated \cite{burke2020gradient}. Such methods are geared towards cases where the evaluated function is a black box. In our differentiable language, we instead take advantage of the fact that the control logic of the program is known and calculate the interpolation \emph{during} the execution. Since discontinuities are caused by comparison statements during the control flow's evaluation, we can derive a \emph{distance measure} from the discontinuity at the evaluation of each control flow statement. As we will see throughout the paper, the advantage of this approach is that it allows an interpolation identical to the case-based view, while avoiding the need to evaluate all paths through the control flow.

After smoothly interpolating discontinuities caused by control flow, we can use the reverse mode of AD \cite{griewank2003mathematical} for derivatives, while retaining the advantage which is otherwise unique to finite differences: A representation of the derivative's trajectory which accounts for discontinuities \cite{more2014you}. Two main challenges arise in this work: First, compositions of functions need special treatment. Second, only some of a given program's control flow paths are relevant to calculate a smooth interpolation. Identifying these paths translates to a combinatorial problem that we attempt to solve.
\\
\\
This paper is structured as follows: In Sections \ref{sec:formalities}-\ref{sec:combinatorial_problem_and_pruning}, we build some theory using the notion of discontinuous functions, and introduce the challenges for smoothing and efficiently evaluating compositions thereof. Sections \ref{sec:implementation}-\ref{sec:algorithmit_smooth_differentiation} introduce an algorithm for efficient smooth evaluation and present its implementation in practical settings. The paper concludes with a case study presenting an application for smooth AD in discontinuous optimization.

%to smoothing generic discontinuities, the ones conducive to our work are those which interpolate between cases, 

%the ones of generic functions either automatically, or which could be read as such. However, those regard the smooth interpolation as mostly an exercise 

\section{Formalities and Basic Method}
\label{sec:formalities}
%TODO: Introduce: We consider logical statements which 
%what's a boundary point? What's a condition $c(x)$? What's a contribution?

We consider a piecewise continuous function $f$, which switches between two subfunctions based on a boundary $\xi$, which (in general) is a manifold serving to delineate a $n$-dimensional space into any two subspaces $S_1, S_2 \subseteq \mathbb{R}^n$, $S_1 \cap S_2 = \emptyset$, $S_1 \cup S_2 = \mathbb{R}^n$. We define $c_{\xi}(x)$ such that it evaluates to \emph{true} exactly if $x \in S_1$.
\begin{equation}
	f(x) = 	
	\begin{cases}\begin{aligned}
			f_1(x), && c_{\xi}(x)\\
			f_2(x), && \overline{c}_{\xi}(x) % \color{gray!75}{\Leftrightarrow x \in S_2}
	\end{aligned}\end{cases}
	:=
	\begin{cases}\begin{aligned}
			f_1(x), && d_{\xi}(x) \leq 0\\
			f_2(x), && d_{\xi}(x) > 0 % \color{gray!75}{\Leftrightarrow x \in S_2}
	\end{aligned}\end{cases}
	\label{eq:switch_idea}
\end{equation}
where the function $d_\xi$ describes the distance from boundary $x$. Thus, $\xi := \{x \mid d_\xi(x) = 0\}$ and describes the distance from it. We furthermore define $\xi$ to be oriented such that $c_\xi$ is true exactly if $d_\xi \leq 0$.

The principle behind our interpolation can be understood as \emph{the probability of choosing another subfunction given a random perturbation} to $d_\xi$. A perturbation will be drawn from a distribution centered at $0$, whose probability decays with distance. Adding a sample from this distribution to the distance function $d_\xi$ is most likely to change the distance only slightly. To formalize this, let the random variable $X \in \mathbb{R}$ be given by a probability distribution $P[X = x] = \sigma'(x)$. Of $\sigma'$ we require that it be symmetric at $0$, continuously differentiable, that $\sigma' > 0$, that $\frac{d}{dx}(\sigma'(x)) \geq 0$ for all $x \leq 0$, and that the integral $\sigma(x) \equiv \int_{-\infty}^{x}\sigma'(x)$ is a probability density-function, i.e. $\int_{-\infty}^{\infty}\sigma' = 1$.

%We draw a $q \sim \sigma'(X)$ and determine whether $q$ falls further from the middle of the distribution than $x$ is from $\xi$ (preserving the sign of the distance). 
To express the perturbation, we draw a $q \sim X$ and add it to the distance compared in the case selector. In this manner, the cases are chosen probabilistically
\begin{equation}
	\hat{f}(x) = 	
	\begin{cases}\begin{aligned}
			f_1(x), && d_{\xi}(x) + q \leq 0 \\
			f_2(x), && d_{\xi}(x) +q > 0
			%f_1(x), && p( d(x,\xi_f) < q ) \\
			%f_2(x), && p( d(x,\xi_f) \geq q )
	\end{aligned}\end{cases}
	\label{eq:stochastic_form}
\end{equation}
%where $d(\cdot, \xi)$ is the signed distance function from manifold $\xi$. 
Cases $f_1$ and $f_2$ are obtained with probabilities $p(d_\xi(x + q) \leq 0)$ and $p(d_\xi(x + q) > 0)$, respectively. For the expectation value $\tilde{f}(x) := E [\hat{f}(x)]$ that means

\begin{equation}
	\begin{aligned}
		\tilde{f}(x) = && E[p(d_{\xi}(x) + q) \leq 0] && \cdot f_1(x) + && E[p(d_{\xi}(x) + q) > 0] && \cdot f_2(x) \\		
		= && \sigma(d_\xi(x)) && \cdot f_1(x) + && \overline{\sigma}(d_\xi(x)) && \cdot f_2(x) \\
		\equiv  && \sigma_{\xi}(x) && \cdot f_1(x) + && \overline{\sigma}_{\xi}(x) && \cdot f_2(x)
	\end{aligned}
	\label{eq:contribution_form}
\end{equation}
where $\sigma$ is the integral of $\sigma'$. Above, we used the definition $\overline{\sigma} \equiv 1- \sigma$. This is useful because $\sigma \in [0,1]$ represents a probability density function, of which $\overline{\sigma}$ is the composite. Taking the expectation value of probabilities from each case of equation (\ref{eq:stochastic_form}) into a case's \emph{contribution}. Contributions are represented by the factors before $f_1$ and $f_2$. Note that these contributions add up to $1$. 

In this work, we will use only piecewise-defined functions with two cases. Functions with an arbitrary number of cases can be obtained by nesting these. Furthermore, binary piecewise-defined functions mimic the way arbitrary hyperplanes are defined in a programming language and also guarantee that all resulting cases constitute a division of the input space. %(i.e. $\mathbb{R}^ n = $).
\label{sec:method}

\section{Composed Functions Require a Specific Smoothing Method}
In an earlier paper, Zang \cite{zang1981discontinuous} introduced an analogous idea differently. In his work, discontinuities are represented as step functions at different boundaries of each spatial subdivision. An interpolation is then computed between \emph{all} cases, based on their boundaries. 

This computation can become extensive. We introduce a method to track boundary distances only if they are needed for the evaluation. This reduces the computational effort of the interpolation considerably. Importantly, we generalize the idea of smooth interpolation to nested functions. This allows us to apply our method to program code.

When nesting piecewise continuous functions, their discontinuities proliferate into the nesting as well. As a result, the induced boundary points are no more known in terms of input space, but implicitly,
\\
\begin{equation}
	\begin{aligned}
		f(x) = 
		\begin{cases}
			f_1 & c_{\xi_f}(x) \\
			f_2 & \overline{c}_{\xi_f}(x)
		\end{cases}, 
		\text{\hspace{0.4cm}}
		g(x) = 
		\begin{cases}
			g_1 & c_{\xi_g}(x) \\
			g_2 & \overline{c}_{\xi_g}(x)
		\end{cases}
		\text{\hspace{0.4cm}% and \hspace{0.4cm}
		}
		\\
		f \circ g =
		\begin{cases}
			f_1 \circ g_1 & c_{\xi_g}(x) \wedge c_{\xi_f}(g_1(x)) \\
			f_1 \circ g_2 & \overline{c}_{\xi_g}(x) \wedge c_{\xi_f}(g_2(x)) \\
			f_2 \circ g_1 & c_{\xi_g}(x) \wedge \overline{c}_{\xi_f}(g_1(x)) \\
			f_2 \circ g_2 & \overline{c}_{\xi_g}(x) \wedge \overline{c}_{\xi_f}(g_1(x)) \\
		\end{cases} \hspace{1.2cm}
	\end{aligned}
	\label{eq:composition_discrete}
\end{equation}
because conditions $c_\xi$ in the cases of $f\circ g$ contain results from functions further upstream. Expressing the cases in equation (\ref{eq:composition_discrete}) explicitly (in terms of the input space) would require function inverses. Since inverses only exist under certain conditions (see for example \cite{arutyunov2010existence}), it is more practical to smoothly interpolate without knowledge of the boundary points in input space.

\subsection{Smooth, then Compose}

\begin{figure}[h!]
	\hspace{-0.7cm}\hbox{\scalebox{0.48}{\input{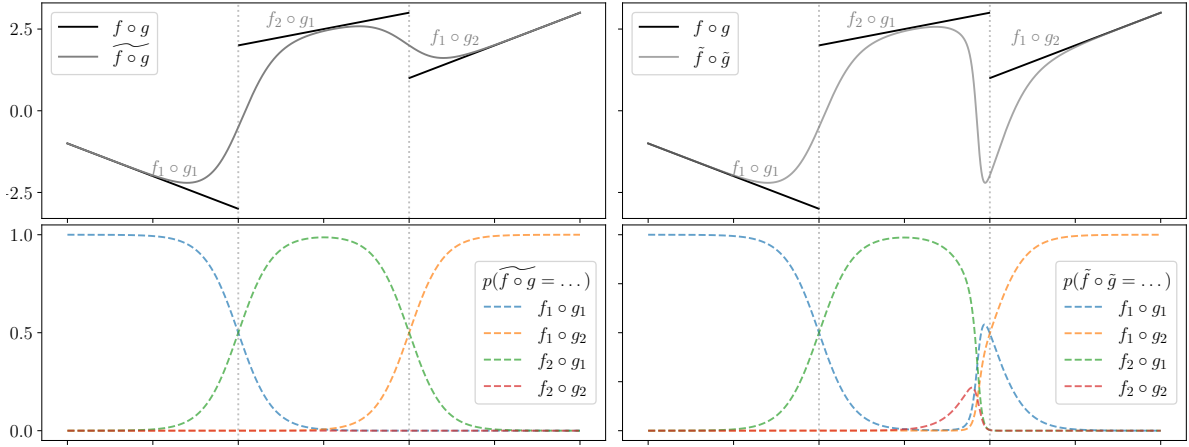}}}
	%\hspace{-0.7cm}\hbox{\scalebox{0.48}{\input{content/figures/artefact_vs_proper.pgf}}}
	\caption{Smoothing approaches and contributions for $\widetilde{f\circ g}$ and $\tilde{f}\circ \tilde{g}$}
	\label{fig:figure_with_contributions}
\end{figure}

Before introducing our method, we point out how simply composing smoothed functions leads to a smooth interpolation with an undesirable trajectory. Consider composing $\tilde{f}$ and $\tilde{g}$ to $\tilde{f} \circ \tilde{g}$. Following the scheme of equation (\ref{eq:contribution_form}), we obtain 
\begin{equation*}
	\begin{aligned}
		\tilde{f} \circ \tilde{g} \mult (x) = & &\sigma_{\xi_f}(\tilde{g}(x)) \cdot f_1( \tilde{g}(x) ) \\
		& +&  \overline{\sigma}_{\xi_f}(\tilde{g}(x)) \cdot f_2( \tilde{g}(x) ) \\
	\end{aligned}
\end{equation*}
where
\begin{equation*}
	\begin{aligned}
		\tilde{g}(x) = \sigma_{\xi_g}(x) \cdot g_1(x) + \overline{\sigma}_{\xi_g}(x) \cdot g_2(x)
		%	\tilde{f} \circ \tilde{g} \mult (x) 
		%	& = &\tilde{f}(&\sigma_{\xi_g}(x) \cdot g_1(x) + \overline{\sigma}_{\xi_g}(x) \cdot g_2(x)&) \\
		%	& = \sigma_{\xi_f}(\dots) \cdot &f_1( &\dots& )  \\
		%	& + \overline{\sigma}_{\xi_f}(\dots) \cdot &f_2( &\dots& )	
	\end{aligned}
	%\kappa_{\xi_f} \cdot f_1(\kappa_{\xi_g} \cdot g_1 + \overline{\sigma}_{\xi_g} \cdot g_2) + \overline{\sigma}_{\xi_f} \cdot f_2(\kappa_{\xi_g} \cdot g_1 + \overline{\sigma}_{\xi_g} \cdot g_2)	
\end{equation*}
We can find the contributions that describe how much each of the cases of the discrete version from equation (\ref{eq:composition_discrete}) account for values of interpolation $\tilde{f} \circ \tilde{g}$. Note that $\sigma_{\xi_g}(x)$ determines the local contribution of $g_1$ %and $g_2$
. Further downstream in the composition, the local contribution of $f_1$ %or $f_2$, 
is determined by $\sigma_{\xi_f}(\tilde{g}(x))$. Overall, in $\tilde{f} \circ \tilde{g}$ the contribution of case $f_1 \circ g_1$ is given by
\begin{equation*}
	\begin{aligned}
		E\left[p(\tilde{f} \circ \tilde{g} = f_1 \circ g_1)\right] = \sigma_{\xi_g}(x) \cdot \sigma_{\xi_f}(\tilde{g}(x)) \\
		%\kappa_{f_1 \circ g_2} = \sigma_{c_f}(\tilde{g}(x)) \cdot \sigma_{\overline{c}_g}(x)	\\
		%\vdots	
	\end{aligned}
\end{equation*}
This does not fit the probabilistic interpretation of $f \circ g$ as it was introduced in section \ref{sec:method}. The probabilistic interpretation requires that a perturbation be applied to each selector ($c_{\xi_f}$ and $c_{\xi_g}$) in eq. \ref{eq:composition_discrete} by perturbing their distance functions. Drawing a $q \sim X$ and $r \sim X$ from random variable $X$ (defined in section \ref{sec:formalities}) to perturb each of the selectors, we obtain the first case $f_1 \circ g_1$ with probability

\begin{equation}
	\begin{aligned}
		p(\widetilde{f\circ g} = f_1 \circ g_1) &= p(d_{\xi_g}(x)+ q \leq 0 \wedge d_{\xi_f}(g_1(x))+ r> 0)  \\
		& = p(d_{\xi_g}(x )+ q\leq 0)  \cdot p(d_{\xi_f}(g_1(x))+ r > 0)
	\end{aligned}
	\label{eq:contribution_f1_o_g1}
\end{equation}
from which we again take the expectation to obtain the contribution $\kappa_{f_1 \circ g_1} := E[p(\widetilde{ f\circ g} = f_1 \circ g_1)]$, meaning
\begin{equation*}
	\begin{aligned}
		\kappa_{f_1 \circ g_1}(x) = \sigma_{\xi_g}(x) \cdot \overline{\sigma}_{\xi_f}(g_1(x))
	\end{aligned}
\end{equation*}
and all other contributions analogously. A comparison between the two smoothing approaches is shown in Figure \ref{fig:figure_with_contributions}. Compared to $\widetilde{f\circ g}$, the trajectory of $\tilde{f} \circ \tilde{g}$ at the transition from cases $f_2 \circ g_1$ to $f_1 \circ g_2$ includes a deviation. It is caused by the fact that contributions that do not stem from either of these two transitioning cases, shortly peak. In $\widetilde{f \circ g}$, the contribution of a case strictly decreases with distance from the case's boundary.

\subsection{Compose, then Smooth}

Our perspective allows to simply interpret each case probabilistically by applying a perturbation in the case selection. One obtains all contributions $\{\kappa_{f_i \circ g_i}\}_{i,j \in \{1,2\}}$ by taking the expected value of the case-selection after perturbation as in equation (\ref{eq:contribution_f1_o_g1}). Weighing all cases of equation (\ref{eq:composition_discrete}) by their contributions, we obtain the interpolation
\begin{equation*}
	\widetilde{f\circ g} := \sum_{i,j \in \{1,2\}} f_i \circ g_j \cdot \kappa_{f_i \circ g_j}
	\label{eq:f_g_smooth}
\end{equation*}
where refer to the each case's weight $\kappa_{f_i \circ g_j}$ as its \emph{contribution}.

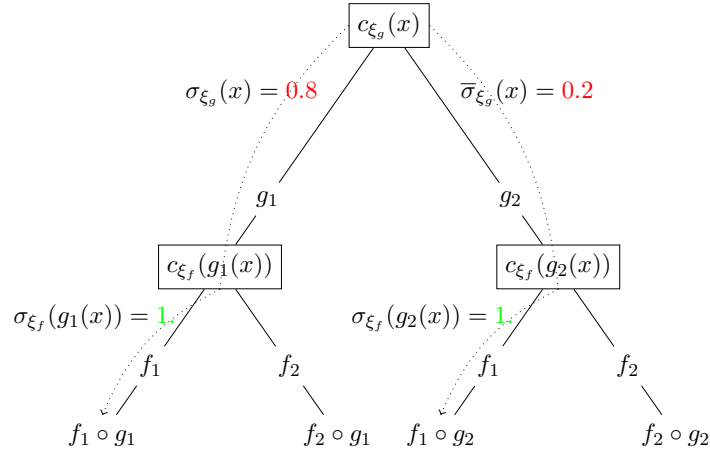
\begin{figure}[h!]
	\centering
	\scalebox{0.86}{\begin{forest}
	for tree={
		calign=fixed edge angles,
		base=bottom{},
		%text height=2ex, 
		%anchor=north,
		%grow=south,
	}
	[$c_{\xi_g}(x)$,draw, name=xg
		[$g_1$, edge label={node[midway,above left]{$\sigma_{\xi_g}(x) = \color{red}0.8$}}
			[$c_{\xi_f}(g_1(x))$, draw, name=xf-left
				[$f_1$, edge label={node[midway, left]{$\sigma_{\xi_f}(g_1(x)) = \color{green}1.$}}
					[$f_1 \circ g_1$, name=f1g1]
					[,phantom]
				]
				[$f_2$
					[,phantom]
					[$f_2 \circ g_1$]
				]
			]
			[,phantom]
		]
		[$g_2$, edge label={node[midway,above right]{$\overline{\sigma}_{\xi_g}(x) = \color{red}0.2$}}
			[,phantom]
			[$c_{\xi_f}(g_2(x))$,draw, name=xf-right
				[$f_1$, edge label={node[midway, left]{$\sigma_{\xi_f}(g_2(x)) = \color{green}1.$}}
					[$f_1 \circ g_2$, name=f1g2]
					[,phantom]
				]
				[$f_2$
					[,phantom]
					[$f_2 \circ g_2$, name=f2g2]%,onslide=<3->{path highlight}]
				]
			]
		]
	]	
	\path[-, dotted, bend right=20] (xg.west) edge (xf-left.south);
	\path[-, dotted, bend left=20] (xg.east) edge (xf-right.south);
	\path[->, dotted, bend right=20] (xf-left.south) edge (f1g1.north);
	\path[->, dotted, bend right=20] (xf-right.south) edge (f1g2.north);
\end{forest}}
	\caption{With pruning, $\widetilde{f\circ g}$ %(eq. \ref{eq:f_g_smooth})
	avoids evaluating the paths with $\kappa_p < \epsilon$. While $\epsilon$ is machine precision here, other values are discussed in Section (\ref{sec:epsilon_h_tradeoff})
}
	\label{fig:tree_tracing}
\end{figure}

We now formalize this idea. We use the operator $\bigcirc$ to define the sequential composition of functions, from the outermost to the innermost function. We consider $k$ piecewise-defined functions  $Q = \underset{i \in 1 \dots k}{\bigcirc} \mult q^{(i)}$, where each $q^{(i)}$ switches between ${q^{(i)}_1}$ and ${q^{(i)}_2}$ at the boundary $\xi_{q^{(i)}}$. All possible cases in $Q$ are represented by paths $\mathcal{P} = \{p \mid p \in \{1,2\}^{\abs{p}} \}$ through the nesting. We use $\abs{p}$ to denote the length of a path. For
\begin{equation}
	q^{(i)} = \begin{cases}
		\begin{aligned}
			{q^{(i)}}_1, && c_{\xi_{q^{(i)}}} \\
			{q^{(i)}}_2, && \overline{c}_{\xi_{q^{(i)}}} 
		\end{aligned}
	\end{cases} 
	\text{\hspace{1cm}we obtain\hspace{1cm}}
	\widetilde{Q} = \sum_{p \in \mathcal{P}}  \kappa_p \cdot \underset{i \in 1\dots \abs{p}}{\bigcirc} q^{(i)}_{p_i} 
	\label{eq:concatenation}
\end{equation}
where a path's contribution $\kappa_p$ is the product of the \emph{local} contributions from each switch along the path
\begin{equation}
	\kappa_p = 
	%\prod_{j \in \{1 \dots |p|\}} \kappa_{p_j} =
	\prod_{j \in \{1 \dots \abs{p}\}} \kappa_{p_j}
	\label{eq:leaf_contribution} 
\end{equation}
and where local contributions depend on the input evaluated in the control-flow statement
\begin{equation*}
	\kappa_{p_j} =
	\begin{cases}
		\sigma_{\xi_{q^{(j)}}} (\underset{j<l\leq \abs{p}}{\bigcirc}  q^{(l)}_{p_l}) & \text{if } c_{\xi_j} \\
		\overline{\sigma}_{\xi_{q^{(j)}}} (\underset{j<l\leq \abs{p}}{\bigcirc}  q^{(l)}_{p_l}) & \text{otherwise}
	\end{cases}	
\end{equation*}
%j<l\leq n
We note that $\kappa_p$ for any path $p$ is made up of $\kappa_{p_j}$ calculated during the composition. From a computational perspective, it is useful to apply the composition up to every $q^{(j)}$, and calculate $\kappa_{p_j}$. This way, all $\kappa_{p_j}$ can be preaccumulated during the path's computation.
This becomes clearer in Figure \ref{fig:tree_tracing}, where we show the downstream flow of input as a tree where all control flow paths are represented. As the paths branch off at every condition, a tree can be used to visualize the downstream flow of the composition's input through the nested functions and conditions. 
\\
\\
The tree resulting from a composition is binary but not necessarily symmetric: at some branchings, one branch may entail a piecewise-defined function, while the other may entail a function defined on the whole domain. Furthermore, paths $p$ in a nesting do not necessarily all have the same length. These points are further illustrated in Section \ref{sec:examples}, where we map generic control flow in code to a tree.

\section{Combinatorial Problem and Pruning}
\label{sec:combinatorial_problem_and_pruning}
The tree corresponding to the paths of $\widetilde{f \circ g}$ is visualized in Figure \ref{fig:tree_tracing}, where each node represents a decision point, and each edge represents a subfunction. It bears evidence to the following combinatorial problem: Each time a branch is encountered, two new subtrees have to be evaluated. Evaluating the smooth interpolation (equation (\ref{eq:concatenation})) means evaluating all leaves of the tree together with their contributions, since mathematically each path contributes with $\kappa_p > 0$. 

However, evaluating all leaves also means to run paths with contribution $\kappa_p \approx 0$. Implementing equations (\ref{eq:concatenation})-(\ref{eq:leaf_contribution}) directly would thus include paths where $\kappa_p$ is zero numerically. That means, these paths would still be executed despite not affecting the result. We want to recognize such cases in order to prune subtrees whose evaluation does not contribute. Such paths are given by the fact that the successive multiplication of local contributions until their $\ell$-th edge causes the product $\kappa_p$ to become zero numerically in equation \ref{eq:leaf_contribution} before the end of the path $\ell \leq \abs{p}$. Formally
\begin{equation}
	\kappa_p = \prod_{j \in \{1 \dots \ell\}} \kappa_{p_j} < \epsilon
	\label{eq:branch_below_machine_precision}
\end{equation}
where $\epsilon$ is chosen as machine precision. We can use equation (\ref{eq:branch_below_machine_precision}) as a guide for a less strict, but simpler heuristic. Instead of keeping track of whether an executed branch is eventually cut, we can directly identify irrelevant subtrees during execution: At every tree node where  $\kappa_{p_j} > 1 - \epsilon$, the opposite branch has a contribution $< \epsilon$ and can be pruned. Branches not pruned by this heuristic, are added to the result and will be referred to as \emph{relevant paths}.

%More intuitively the above formula says that on a path $p$, we encountered the $\ell$-th condition and see that one of its subtrees will have a contribution of less than machine precision, ergo $0$. Such subtrees can be pruned, since they do not influence the result. The remaining paths with $\kappa_p \geq \epsilon$ we call \emph{relevant paths}. 

\subsection{Smoothing Sharpness}

Until here it was assumed that the size of the perturbation is fixed. We introduce a parameter $h \in \mathbb{R}$ to describe the range of the perturbation. Mathematically, we expand re-purpose the function $\sigma'$ given in Section \ref{sec:method} to describe a random variable $Y$ given by the probability distribution $P[Y = x] = \frac{1}{h} \sigma'(h\mult x)$, where $h$ describes the \emph{intensity} of the perturbation. Again, we sample $q  \sim Y$ and $r \sim Y$. Analogous to equation (\ref{eq:contribution_form}), we obtain

\begin{equation}
	\begin{aligned}
		\tilde{f}(x) = && E[p(d_{\xi}(x) + q \leq 0)] && \cdot f_1(x) + && E[p(d_{\xi}(x) + r > 0)] && \cdot f_2(x) \\		
		= && \sigma\left( d_\xi(x) \cdot h \right) && \cdot f_1(x) + && \overline{\sigma} \left( d_\xi(x) \cdot h \right) && \cdot f_2(x) \\
		\equiv  && \sigma_{\xi, h}(x) && \cdot f_1(x) + && \overline{\sigma}_{\xi, h}(x) && \cdot f_2(x)
	\end{aligned}
\end{equation} 
\\
Here, decreasing $h$ reduces the sharpness of the transition, and increasing $h$ makes transitions sharper. In the extreme, $h \to \infty$ leads to the discrete formulation of equation (\ref{eq:switch_idea}), because $\underset{h \to \infty}{\sigma(h \cdot x)}$ is either $0$ or $1$, depending on the sign of $x$.

Furthermore, $h$ has an effect on the combinatorial problem: It changes the number relevant paths with $\kappa_p \geq \epsilon$. This is because for each factor $\sigma_{\xi_{q^{(i)}}}$ in equation (\ref{eq:branch_below_machine_precision}), it holds that $\sigma_{\xi_{q^{(i)}}, h_1}(x) \geq \sigma_{\xi_{q^{(i)}}, h_2}(x)$ for all $x$ if $h_2 > h_1$. Overall, the higher $h$, the more branches we are likely to require evaluation for any given input.

\subsection{Precision and Combinatorial Problem}
\label{sec:epsilon_h_tradeoff}
Datatypes differ on machine precision. Accordingly, a datatype's precision gives the natural choice for $\epsilon$ (in equation (\ref{eq:branch_below_machine_precision})). Intuitively speaking, the distance function defines a range $\{x \mid 1-\epsilon \geq \sigma(d_\xi(x)) \geq \epsilon\}$  centered at $\xi$, within which the interpolation evaluates both subtrees of condition $c_\xi$. Beyond the range, only one of the subtrees is evaluated. The resulting interpolation is smooth as long as $\epsilon$ is chosen to be machine precision (or lower) because numerically, testing \verb"<" $\bm{\epsilon}$ on a machine is equivalent to comparison to zero. which computes all relevant paths in the tree computed.

%the interpolation calculates all sets of relevant paths for any point. As a result, the interpolated function remains smooth between any inputs which lead to different sets of relevant paths.
On the contrary, setting $\epsilon$ higher than machine precision means that at certain nodes, subtrees may be pruned despite containing a relevant path to the sum in equation \ref{eq:concatenation}. The resulting function is not guaranteed to be continuous at points where a small change in input leads to a change in the set of relevant paths. This is because some paths with $\kappa_p > 0$ can be absent in the sum of equation \ref{eq:concatenation}.

Nonetheless, it comes with a trade-off: Pruning branches with low contributions $\kappa_p < \epsilon$ reduces the size of the combinatorial problem. This allows to contain the effect of the sharpness parameter $h$ on the problem size: One can lower the interpolation's sharpness with an arbitrary $h$ while separately controlling the number of evaluated paths through $\epsilon$.  %one may choose arbitrarily sharp or smooth transitions, while reducing the number of evaluated paths.

\section{Notes on the Implementation}
\label{sec:implementation}
\subsection{Probabilistic Interpretation of Conditions}
\label{sec:probabilistic_conditions_and_clauses}
To be applicable to code, the probabilistic defined in our method must map to boolean logic. In Boolean logic, one can aggregate conditions to clauses via logical connectives. To define a clause's probabilistic interpretation, we consider the connectives $\{\wedge, \lor, \overline{\text{\phantom{a}}}\}$ which are present in programming languages.
Consider the conditions $c_{\xi_1} := d_{\xi_1} \leq 0$ and $c_{\xi_2} := d_{\xi_2} \leq 0$. Their analogous expressions in probabilistic logic are given by taking the expected probabilities given a perturbation with samples $q, r$ as in Section \ref{sec:method}.
\begin{table}[h!]
	\tbl{Probabilistic Equivalents to Boolean Conditions given a Perturbation}{
		\def\arraystretch{1.5}
	\begin{tabular}[l, scriptsize=\tiny]{ c | l }		
		Boolean & Probabilistic \\ \hline	
		$c_{\xi_1} \wedge c_{\xi_2}$ & 
			
				$ \phantom{= } E[p(d_{\xi_1}(x) + q\leq 0 \wedge d_{\xi_2}(x) + r \leq 0)]$ \\
				&$ = E[p(d_{\xi_1}(x) + q\leq 0)] \cdot E[p(d_{\xi_2}(x) + r \leq 0)]$ \\
				& $= \sigma_{\xi_1}(x) \cdot \sigma_{\xi_2}(x)$
			\\ \hline
		
		$c_{\xi_1} \lor c_{\xi_2}$ & 
			$ \phantom{=} E[p(d_{\xi_1}(x) + q\leq 0 \lor d_{\xi_2}(x) + r \leq 0)] $ \\
			& $= E[p(d_{\xi_1}(x) + q\leq 0)] \cdot E[p(d_{\xi_2}(x) + r \leq 0)] - E[p(d_{\xi_1}(x) + q\leq 0)] \cdot E[p(d_{\xi_2}(x) + r \leq 0)] $ \\
			& $= \sigma_{\xi_1}(x) + \sigma_{\xi_2}(x)] - \sigma_{\xi_1}(x) \cdot \sigma_{\xi_2}(x) $
			\\ \hline
		
		$\overline{c}_{\xi_1}$ & $E[1 - p( d_{\xi_1}(x) + q \leq 0)] = 1 - \sigma_{\xi_1}(x)$ \\
		 %\(\displaystyle \sum_{n=1}\nolimits' C_n \)  & a
 	\end{tabular}}

	\label{tab:tendential_logic}
\end{table}
The definitions in Table \ref{tab:tendential_logic} also correspond with the interpretation of Zadeh's fuzzy logic \cite{zadeh1988fuzzy}.

Next we turn to comparison operators $\{<, \leq, \geq, >\}$. Note that probabilities for $\{\leq, >\}$ have been defined in equations \ref{eq:switch_idea}-\ref{eq:contribution_form} as $\sigma(d_\xi(x))$ and $\overline{\sigma}(d_\xi(x))$. Due to the point symmetry of $\sigma$, we see that
\begin{equation}
	\begin{aligned}
		& \text{ } & \sigma(d_\xi(x)) &=& \overline{\sigma}(&-d_\xi(x))& \\
		\Leftrightarrow  && E[p(d_\xi(x) \leq 0)] &=& E[p( - &d_\xi(x) > 0)]& = E[p(d_\xi(x)  < 0)]\\
		% &&& = & p(d_\xi(x) < 0)
	\end{aligned}
	\label{eq:operator_leq_is_le}
\end{equation} 
In other words, contributions of $\{<, \leq\}$ are represented by the same function in our probabilistic logic. Analogously, Contributions of $\{>, \geq \}$ are represented by the formers' composite. More concretely, this means that any $c_\xi$ which is true at $d_\xi \leq 0$, has its expected value represented by a function $\sigma(d_\xi(x))$, that also represents $d_\xi < 0$.
\\
We have yet to treat the equals $(=)$ operator of the probabilistic logic. It can be constructed from the previously defined operators
\begin{equation*}
	\begin{aligned}[c]
		 &p(x = \xi)&=\mult\mult& p(d_\xi(x) \leq \xi \wedge d_\xi(x) \geq \xi) = \mult p(d_\xi(x) \leq \xi) \cdot p(d_\xi(x) \geq \xi) \\
		\Leftrightarrow \mult E[&p(x = \xi)]& = \mult\mult& \sigma_\xi(x) \mult \cdot\mult  \overline{\sigma}_\xi(x)
	\end{aligned} 
\end{equation*}
With this, all comparison operators and logical connectives are defined in probabilistic logic in analogy to Boolean logic.

\subsection{Control Flow} 
\label{sec:examples}
\begin{figure}[h!]
	\begin{minipage}{0.32\textwidth}
\begin{lstlisting}[language=C++, escapeinside=||, basicstyle=\ttfamily\footnotesize]
  // ...
  if(|$c_{\xi_1}$|(x)){
    x = a(x);
    if(|$c_{\xi_2}(x)$| && |$c_{\xi_3}$|(x))
      x = b(x);
    else 
      x = h(x);
  }
  x = d(x);
  if(|$c_{\xi_4}$|(x))
    x = e(x);
  // ...
\end{lstlisting}
	\end{minipage}
	\begin{minipage}{0.5\textwidth}
		\scalebox{0.7}{\forestset{qtree/.style={for tree={calign=fixed edge angles, base=bottom{}, 
			align=center,inner sep=1pt}}}

\newcommand{\col}[1]{
	{\color{gray!80}#1}}

\begin{forest}, qtree
%	for tree={
%		calign=fixed edge angles,
%		base=bottom{},
%	}
	[$c_{\xi_1}(x)$ \\ \col{$[c_1]$},draw, name=xg
		[$a$ \\ \col{$[f_{1,T}]$}
			[$c_{\xi_2}(a(x)) \wedge c_{\xi_3}(a(x))$ \\ \col{$[c_2]$}, draw, name=xf-left
				[$d \circ b$ \\ \col{$[f_{2,T}]$} ,edge
					[$c_{\xi_4}(d\circ b \circ a (x))$ \\ \col{$[c_4]$} ,draw, name=aa, edge
						[$f$ \\ \col{$[f_{4,T}]$}, edge
							[$f\circ d \circ b \circ a $]
							[,phantom]
						]
						[\phantom{id} \\ \col{$[f_{4,F}]$}, edge
							[,phantom]
							[$d\circ b \circ a$]
						]
					]
					[,phantom]
				]
				[$d \circ h$ \\ \col{$[f_{2,F}]$}
					[,phantom]
					[$c_{\xi_4}(d\circ h \circ a (x))$ \\ \col{$[c_5]$},draw, name=qa, edge
						[$e$ \\ \col{$[f_{5,T}]$}, edge
							[$d\circ h \circ e$]
							[,phantom]
						]
						[\phantom{id }\\ \col{$[f_{5,F}]$}, edge
							[,phantom]
							[$d\circ h$]
						]
					]
				]
			]
			[,phantom]
		]
		[$d$ \\ \col{$[f_{1,F}]$}
			[,phantom]
			[$c_{\xi_4}(d(x))$ \\ \col{$[c_3]$},draw, name=xf-right
				[$e$ \\ \col{$[f_{3,T}]$}, edge %label={node[midway,left]{$\sigma_{x<x_f}(g_2(x))}
					[$d \circ e$, name=f1g2]
					[,phantom]
				]
				[\phantom{id} \\ \col{$[f_{3,F}]$}
					[,phantom]
					[$d$,name=f2g2]
				]
			]
		]
	]
%\path[->, dotted, bend left=20] (xf-right.south) edge (f2g2.north);
\end{forest}

%\begin{forest}, qtree
%	[IP 
%	[Spec]
%	[I\textprime,
%	[I\\are]
%	[VP]
%	]]
%\end{forest}
}	
	\end{minipage}

	\caption{Code (left) and respective control-flow tree (right). Empty edges signify that there is no function changing the input. Control flow can easily become involved. For clarity, we identify functions based on their conditions (gray)}
	
	\label{fig:code_to_tree}
\end{figure}
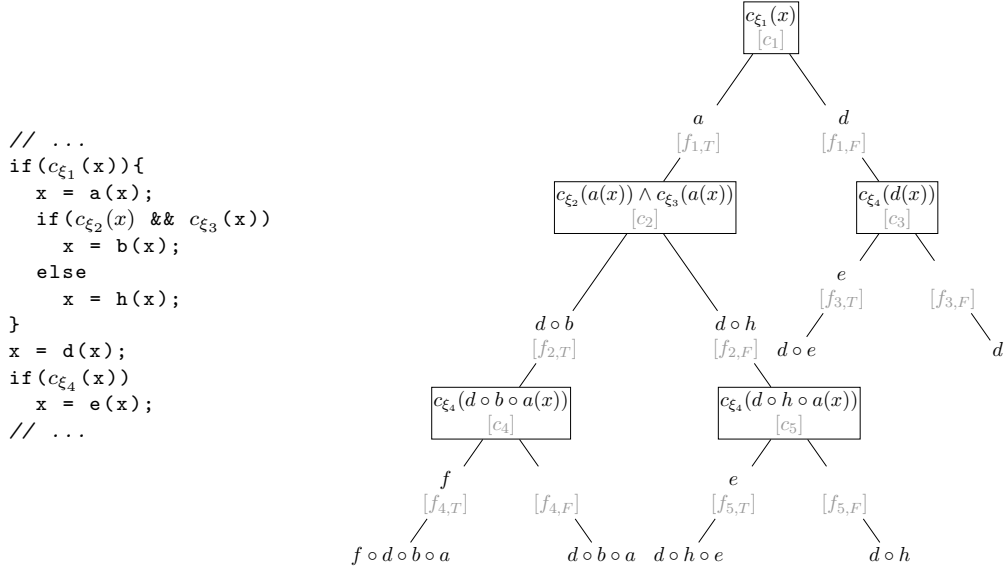

For the remainder of this paper, we simplify the notation for the control flow trees by identifying functions on the branches by their parent conditions: A branch at condition $c_{\bm{i}}$ selects between instruction sequences $f_{{\bm{i}},T}$ and $f_{{\bm{i}},F}$, when evaluating to \verb"true" or \verb"false" respectively. 
This scheme, shown in Figure \ref{fig:code_to_tree}, is more apt at representing any type of control flow and simplifies the notation, by not explicitly naming the instruction sequences inside control flow.
\\
The following Algorithm in Section \ref{sec:tree_tracing} uses the idea of \emph{marking} tree nodes where the interpolation requires both branches to be evaluated. Note that there is no need for the algorithm to ever construct the control flow tree explicitly, or to even be aware of its structure: Markers are set at runtime on certain encountered conditions. The subsequent path execution relies on these markers' information about how to evaluate each condition in the next run. This allows a depth-first traversal restricted to the set of relevant paths.

\subsection{Tree Tracing Algorithm}
\label{sec:tree_tracing}
We regard the function execution as paths through the execution tree given by its control flow (cf. Figure \ref{fig:code_to_tree}). During execution, some conditions $c_{\xi_i}$ will evaluate to \verb"true" with local contribution $\sigma_{\xi_i} \geq \epsilon$, or to \verb"false" with local contribution $\overline{\sigma}_{\xi_i} \geq \epsilon$. These cases indicate that the subtree of the opposite Boolean value at node $c_{\xi_i}$ evaluated along the current path, contains relevant paths. We indicate this by setting markers at such conditions during execution. 

Each path $p$ has a contribution $\kappa_p$, which is calculated by multiplying all local contributions along the path. The procedure of evaluating a single path is shown in Algorithm \ref{alg:if_marked_evaluation}. %New contributing paths (eqs. \ref{eq:concatenation} - \ref{eq:branch_below_machine_precision}), are indicated by markers during the path-execution.
\algdef{SE}[DOWHILE]{Do}{doWhile}{\algorithmicdo}[1]{\algorithmicwhile\ #1}%
\begin{algorithm}[h!]
	\caption{Evaluating control flow and local contributions with markers}\label{alg:if_marked_evaluation}
	\begin{algorithmic}
		%${\boldsymbol f}, 
		\Function{evaluate\_path}{markers}
		\State j $\gets$ 0
		\State $\kappa_p$ $\gets$ 1
		\While{$j$ exists} 
		\State dis $\gets$ $c_{\xi_{\bm{j}}}(x)$ \Comment{discrete value} 
		\State con $\gets$ $\sigma_{\xi_{\bm{j}}}(x)$ \Comment{probabilistic value}
		
		\If{$j \in $ markers $\wedge$ markers[$j$] = false} \State dis $\gets$ $\overline{\text{dis}}$ 
		\State con $\gets$ $1 - $ con
		\ElsIf{$con \geq \epsilon$} markers[$j$] = true 
		\EndIf
		\State $\kappa_p$ $\gets$ $\kappa_p$ * con	\Comment{accumulate contribution}
		\If{dis = true} \Comment{select executed branch}
			\State ${\boldsymbol f}$ $\gets$ ${\boldsymbol f}_{{\bm{j}},T}$
		\Else 
			\State ${\boldsymbol f}$ $\gets$ ${\boldsymbol f}_{{\bm{j}},F}$ 
		\EndIf
		
		\State x $\gets$ ${\boldsymbol f}($x$)$ 
		%\If{!($j \in$ markers) \&\& $\epsilon < con < 1-\epsilon$}
		%	\State markers.[$j$] = true \Comment{Set new marker if interpolation necessary}
		%\EndIf
		\State $j$ $\gets$ next\_condition($j$, dis) \Comment{choose next edge based on discrete Boolean}
		\EndWhile
		\State \Return x, $\kappa_p$, markers
		\EndFunction
	\end{algorithmic}
\end{algorithm}

Once a path is executed, its result is weighted by its contribution and added to the overall result. Afterwards, the most recently set marker (deepest marker) is \emph{negated} by setting its value to \verb"false", unless it has already been negated. Whenever the path execution passes through a negated marker, it forces the Boolean value of the branch and the local contribution to evaluate to their opposites. In other words, negated markers force the control flow of subsequent executions to go down the opposite subtree.

\begin{algorithm}[h!]
	\caption{Tree Tracing}\label{alg:tree-tracing}
	\begin{algorithmic}
		\State $y \gets 0$
		\State markers $\gets$ map\{int, bool\}
		%\Do
		\While{true}
		\State $y_p$, $\kappa_p$, markers $\gets$ \Call{EVALUATE\_PATH}{markers}
		\State $y$ $\gets$ $y + y_p \cdot \kappa_p$
			
		\While{markers.size $> 0 \mult \wedge$ markers.last = false}
			\State markers.remove\_last() \Comment{remove deepest sequence of negated markers}
		\EndWhile

		\If{markers.size $> 0$} 
		\State markers.end $\gets$ false
		\Else \text{ } break
		\EndIf
		\EndWhile
		%\doWhile markers.size $> 0$
	\end{algorithmic}
\end{algorithm}
% If the most deepest marker in the path is traversed after having been negated, we remove this marker and negate the latest remaining marker.
 
If the deepest marker in the tree marker is a negated marker, and the execution passes it without adding a deeper marker, it can be safely removed: This means that all relevant branches in both its subtrees have been executed. The same goes for a consecutive sequence of negative markers that ends in this deepest marker. Overall, the algorithm is akin to a depth-first search of relevant paths. Once all markers have been removed, all relevant paths have been evaluated and added to $y$, weighted by their respective contributions $\kappa_p$. This completes the calculation of equation (\ref{eq:concatenation}) for relevant paths. 
\section{Smooth Algorithmic Differentiation}
\label{sec:algorithmit_smooth_differentiation}
To obtain smooth derivatives, smooth interpolation is used together with AD. Our method of choice for the implementation of the tree tracing (Algorithm \ref{alg:if_marked_evaluation} - \ref{alg:tree-tracing}) is operator overloading in \verb"C++". The overloaded numeric datatype \verb"SType" additionally calculates the local contribution at logical comparators and connectives (cf. Section \ref{sec:probabilistic_conditions_and_clauses}). Their evaluation then triggers the marking process at conditions where a tree node has to evaluate both subtrees.
%\begin{listing}[h!]
%\begin{minted}[fontsize=\normalsize, bgcolor=lightgray!20]{C++}
\begin{lstlisting}[language=C++,  basicstyle=\ttfamily\footnotesize, caption={Schematic Illustration of Operator Overloading with AD Type nested in the Smooth Type}, captionpos=b, label=listing:example, float=h]
 #include <ad_tool>
 #include "smoothing.h"
  
 template<typename T> f(T x1, T x2){ 
   T r = 2;
   
   if (x1*x1 + x2*x2 < 2.) r = r - 1;

   if (x1 < x2) r = r - 1;
   
   return r;
 }

 int main(){
   SType<ADType<double>> x = 2;
   ADType<double> y = run_all_paths(f, x); // run paths until markers consumed
   y.derivative = 1;	                   // seed derivative on smoothed result
   interpret_adjoint(y);                   // adjoint tape accumulation
   double dy_dx = x.derivative;            // harvest smooth derivative
 }
\end{lstlisting}
%\end{minted}
%\end{listing}
An implementation of smooth AD is given in {Listing \ref{listing:example}}. Our implementation available on github\footnote{https://github.com/sebastianfchr/Smoothing\_Essential} uses \verb"dco/c++" \cite{lotz2016hybrid} for the overloaded \verb"ADType". Note that the overloaded \verb"SType" for the smoothing procedure combines easily with any \verb"ADType". This allows for the derivative logic to be run on top of the smoothing logic. Doing so yields the derivatives of the smoothly interpolated function. It is worth noting that we could conversely obtain a smooth interpolation of the derivatives (instead of a derivative of the smooth interpolation) by nesting our types in reverse order to Listing \ref{listing:example}. Although interpolating derivatives is a widespread approach in nonsmooth optimization \cite{burke2020gradient}, we did not further explore this facet of Smooth Algorithmic Differentiation here, because interpolated derivatives discard the information about discontinuities in the function.

\section{A Case Study in Optimization}

To apply our method to a tangible use case, we run a gradient-based descent method on nondifferentiable functions. Gradient-based optimizers find a local minimum of function $f$, starting from any point $x_0$, and essentially progress by following the negative gradient direction
\begin{equation*}
	x_{i+1} = x_i - \ell \cdot \nabla f(x_i)
\end{equation*}
where $\ell$ is the step-size, also called \emph{learning rate}. We consider a function from a collection \cite{lukvsan2000test} of nonsmooth optimization problems: The \emph{crescent} function, given as
\begin{equation*}
	cr(x_1, x_2) = max\{ x_1^2 + (x_2-1)^2 + x_2-1, -x_1^2-(x_2-1)^2 + x_2 + 1\}
\end{equation*}
can be easily formulated as code with one if-statement and then run on the overloaded \verb"SType" for smoothing.
\begin{figure}[h!]
	\centering
	\scalebox{0.55}{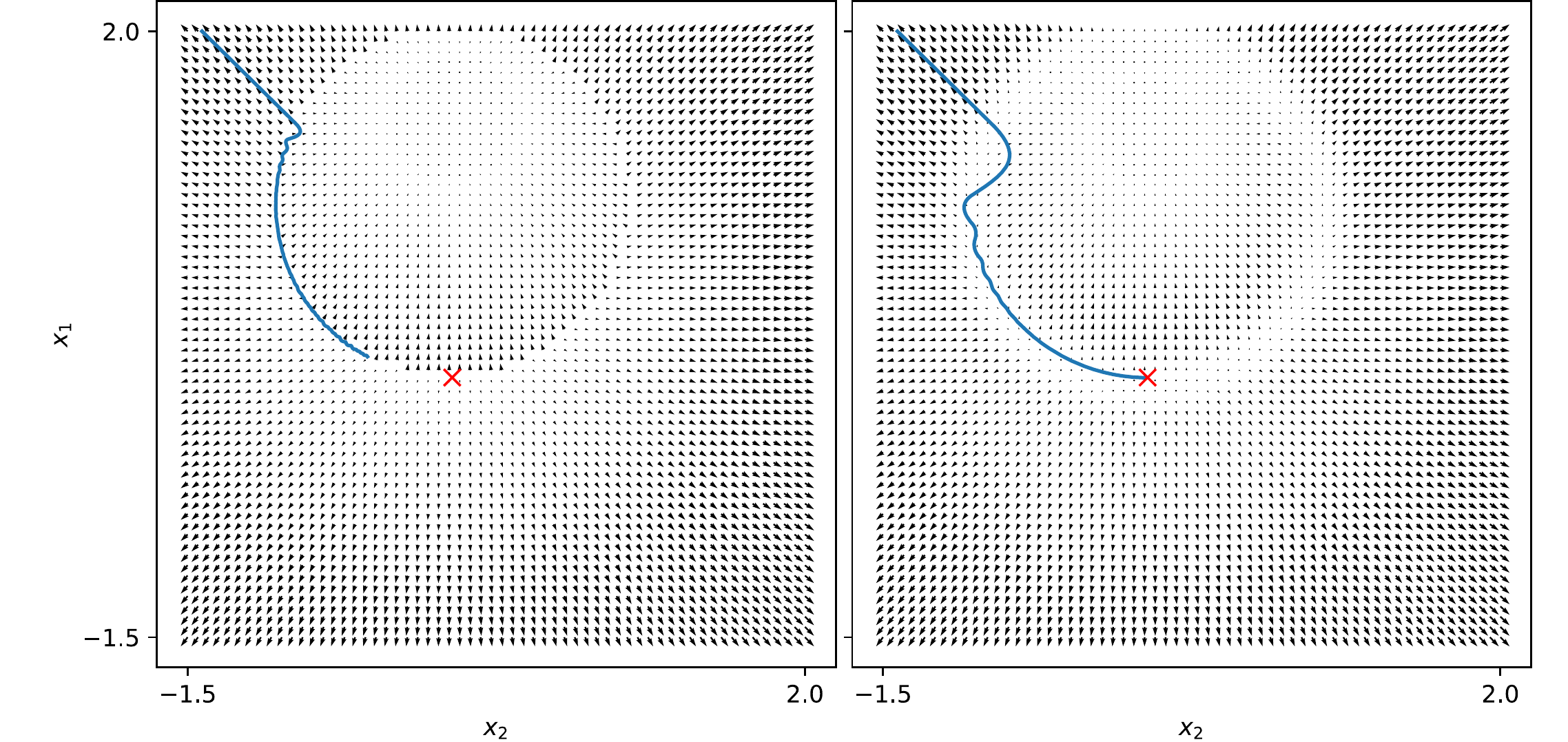} 
	%\scalebox{0.55}{\input{content/figures/smooth_grad_360_marked_onlygrad_BW.pdf_tex}}
	
	\caption{\emph{Crescent} function and gradient field (left column nonsmooth, right column smoothed). The optimizer progressed quicker with the smooth gradient. The line represents the optimizer's trajectory at iteration 300 with sharpness $h = 100$. For visual distinction of the plots, the plotted sharpness $h \ll 100$. }
	\label{fig:smooth_with_gradient_and optimizer}
\end{figure}
The challenge for an optimizer in the above functions is to navigate along the nondifferentiability introduced by the $max$ operator, where gradients cause the optimizer to oscillate back and forth, thus slowing its progress.
\begin{table}[h!]
	\tbl{Optimization outcomes for Crescent per $h$ and number of iterations. (Initial learning-rate = 0.02)}{
		\def\arraystretch{1.5}
		\begin{tabular}[l, scriptsize=\tiny]{c|  c | c | c | c | c | c | c | c}
			\backslashbox{h}{steps} & 200 & 300 & 400 & 500 & 750 & 1000 & 1500 & 2000 \\ \hline 
			10 &0.7116 & 0.0164 & 0.013 & 0.013 & 0.013 & 0.013 & 0.013 & 0.013 \\ \hline
			50 &0.0026 & 0.0026 & 0.0026 & 0.0026 & 0.0026 & 0.0026 & 0.0026 & 0.0026\\ \hline
			100 &0.0013 & 0.0013 & 0.0013 & 0.0013 & 0.0013 & 0.0013 & 0.0013 & 0.0013\\ \hline
			500 &0.0083 & 0.0068 & 0.011 & 0.0012 & 0.0006 & 0.0005 & 0.0006 & 0.0005\\ \hline
			1000 &0.0054 & 0.0109 & 0.0051 & 0.0029 & 0.0005 & 0.0025 & 0.0003 & 0.0011\\ \hline
			$\infty$ & 0.0075 & 0.007 & 0.0086 & 0.0055 & 0.0061 & 0.0035 & 0.0016 & 0.0056\\ \hline
	\end{tabular}}
	
	\label{tab:crescent_smoth}
\end{table}
We hypothesized that smoothing would improve the quality of the minimum or the number of steps needed by the optimizer. We integrated our smoothed functions into TensorFlow \cite{abadi2016tensorflow} to be able to run off-the-shelf optimizers on it, and chose TensorFlow's implementation of the ADAM \cite{reddi2019convergence} optimizer. Our results for optimization with different sharpness-factors $h$ are given in Table \ref{tab:crescent_smoth}. Note that $h=\infty$ translates to the non-smooth original version. We used the recommended starting point for optimizing \emph{crescent} from \cite{lukvsan2000test} and ran the optimizer for various numbers of steps.
\\
For any iteration count, the optimizer could always reach a better minimum for most of the smoothed cases ($h \neq \infty$), compared to the nonsmooth case. Figure \ref{fig:smooth_with_gradient_and optimizer} features a comparison of optimizer's trajectories between a smoothed and non-smoothed case.%the smoothed version progressed more quickly toward the minimum in the \emph{crescent} function.
\\
\\
A more powerful case for our method is the optimization of a piecewise discontinuous function, where discontinuities are not represented in the gradient field with standard algorithmic differentiation. Such an example might look like
\begin{equation}
	g(x_1, x_2) = 
	\begin{cases}
		\begin{aligned}[b]
			0 &\text{\hspace{1cm}} \text{if } {x_1}^2+{x_2}^2 < 2 \wedge x_1 < x_2 \\
			2 & \text{\hspace{1cm}}\text{if } {x_1}^2+{x_2}^2 \geq 2 \wedge x_1 \geq x_2 \\
			1 & \text{\hspace{1cm}}\text{else}  \\ 
			
		\end{aligned}
	\end{cases}
	\label{eq:discontinuous_function}
\end{equation}
\begin{figure}[h!]
	\centering
	\hspace{-0.7cm}\scalebox{0.62}{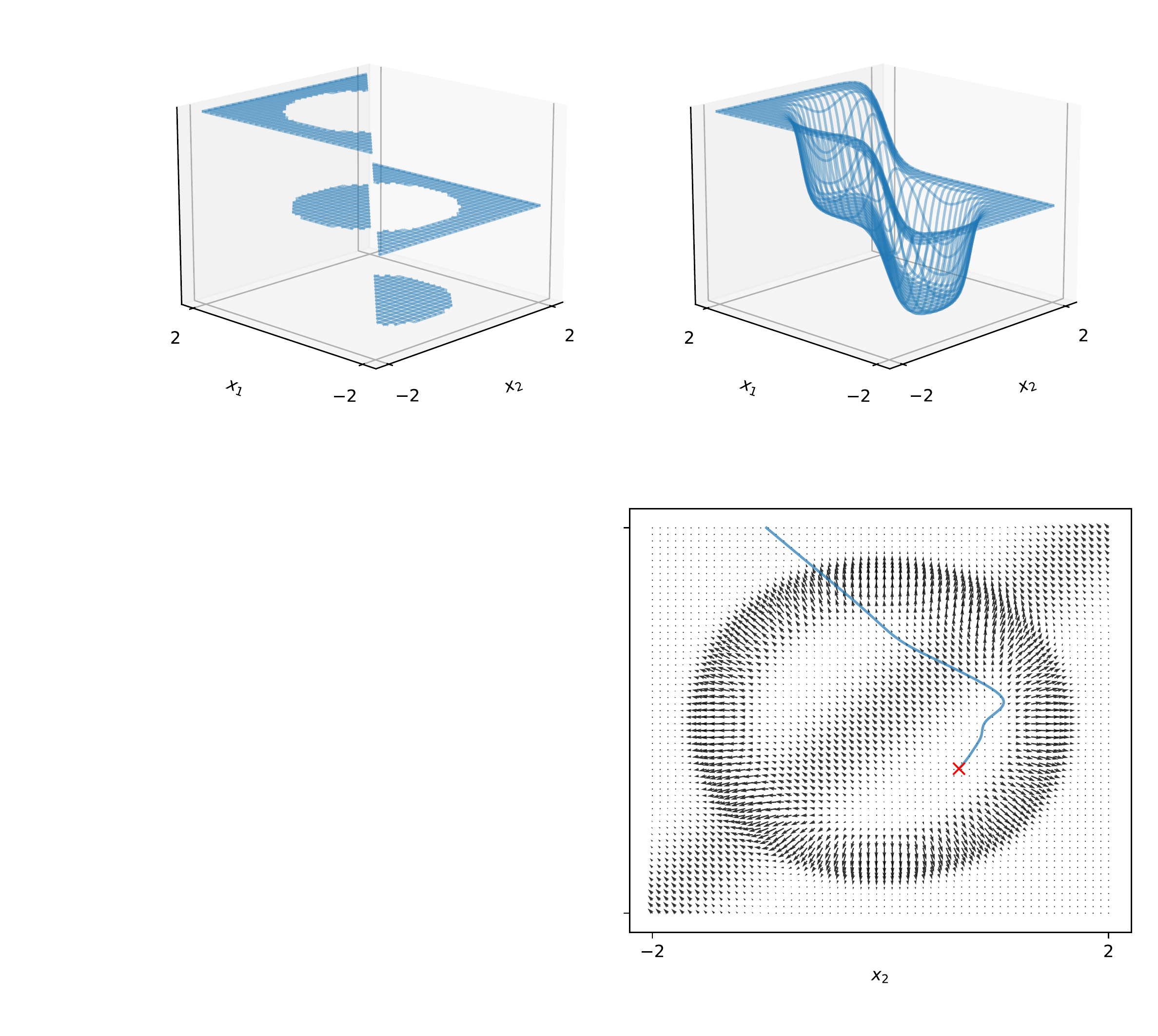}
	%\hspace{-0.7cm}\scalebox{0.62}{\input{content/figures/discontinuous_opt_func_and_grad_BW.pdf_tex}}
	\caption{\emph{Gradients in Discontinuous Functions}. The top left plot shows equation (\ref{eq:discontinuous_function}), which is also expressed as code in Listing \ref{listing:example}. When the smoothed version (top right) is used for taking derivatives, former discontinuities become represented. The gradient field (bottom right) contains the trajectory of an optimizer.}
	\label{fig:discontinuous_with_grad}
\end{figure}
Programming languages can be used to express each of the cases of the above equation (\ref{eq:discontinuous_function}) explicitly, or may also use a more concise formulation as we did in Listing \ref{listing:example}. There, function \verb"f" expresses the exact same behavior. By leveraging our method, we can differentiate a smooth version of its program code. The derivatives taken from the smooth version then represent the sudden change of function values, as can be seen in Figure \ref{fig:discontinuous_with_grad}. As a result, an optimizer can find the minimum of the function.

Our example combines smoothing and optimization, but leaves the sharpness $h$ of the smoothing inflexible. Although useful for illustration, this is by no means the ideal way to harness smoothing in an optimizer. Many optimizers are designed to represent the recent descent history in their parameters and adapt these accordingly \cite{reddi2019convergence}. A yet more promising integration of smoothing and optimization could treat sharpness as a parameter and adapt it in tandem with parameters such as momentum and learning rate.

\section{Outlook}
We introduced a method to interpolate discontinuities caused by control flow without explicit information about the discontinuities in input space. Compared to existing approaches, our method allows for the nesting of functions and interpolates by evaluating only cases whose contribution is relevant. This allows programming in a language where control flow discontinuities are interpolated automatically.

The introduced approach is meant to be integrable into different settings. We took advantage of this to run an out-of-the-box optimizer in discontinuous landscapes and their smoothed counterparts. Despite the fact that our limited case study kept the functionality of optimizer and smoothing separate, it led to improved results.

The smoothing method's execution time is influenced by the smoothed program's control flow, the kernel ($\sigma$) and the parameters for precision ($\epsilon$) and sharpness ($h$), as these affect the number of evaluated paths. Beyond these, execution times of paths through the control flow vary. As a result, runtime is a highly individual issue, which is left up to future research. The core contribution of our idea is to smoothly interpolate program code in an efficient way by evaluating only necessary paths through the control flow. Along this line of thought, is also worth noting that the number (or dimension) of inputs and outputs do not affect the complexity of the algorithm.

We create smooth derivatives by coupling the smoothed program execution with reverse-mode AD. There is ongoing effort to efficiently implement reverse-mode AD in systems capable of massive parallelism \cite{kaler2021parad, van2018automatic}, namely machine-learning and scientific computing. In the future, a parallelized implementation of smoothing will have to follow to be integrable with these fields.

\newpage

%\section{References}
%\cite[cf.]{} => [cf. 8]
%\cite[see][and references therein]{} => [cf. 8 and references therein]

%To produce the list of references, the bibliographic data about each reference item should be listed in \texttt{thebibliography} environment in alphabetical order. References with the same author or group of authors are further sorted chronologically, beginning with the earliest. The following list shows some sample references prepared in Taylor \& Francis' Reference Style S.
%\bibliographystyle{plain}
\newpage
\bibliographystyle{tfs.bst}
\bibliography{content/bibliography}

\section*{Acknowledgements}
This work was partly funded by Numerical Algorithms Group Ltd., Oxford, UK.

\end{document}